\documentclass[aps,prd,twocolumn,amsmath,amssymb,10pt,showpacs]{revtex4-1}
\usepackage{graphicx}

\usepackage{color}

\begin{document}

\title{Long-lived resonances at mirrors.} 

\author{Friedemann~Queisser$^{1}$}
%

\author{William~G.~Unruh$^{2}$}

\affiliation{$^1$Fakult\"at f\"ur Physik, Universit\"at Duisburg-Essen, 
Lotharstrasse 1, 47057 Duisburg, Germany}
\affiliation{$^2$Department of Physics, University of British Columbia, Vancouver, 
V6T 1Z1 Canada}

\date{\today}

\begin{abstract}

Motivated by realistic scattering processes of composite systems, 
we study the dynamics of a two-particle bound system which is scattered at a mirror.
We consider two different scenarios:
In the first case we assume that only one particle interacts directly 
with the mirror whereas in the second case both particles are scattered. 
The coherence between the transmitted and the reflected 
wave-packet is reduced when the internal degree of freedom (the relative coordinate) 
of the bound system becomes excited.
Depending on the particular system-mirror interaction,
long-lived resonances can occur.
%


\end{abstract}


\maketitle

\section{Introduction.} 

The superposition principle is of prime
importance in quantum mechanics and has been demonstrated for 
neutrons \cite{SBRKZ83}, 
in superconductors \cite{RHL95,CCDEM88,SPRR97,NPT99}, 
in nanomagnets \cite{FSTZ96,B99,W97}, with trapped ions \cite{MMKW96}, with photons 
in cavities \cite{B96} and for large molecules \cite{ANVKKZZ99}.
Nevertheless, interferences of macroscopic objects are not yet accessible \cite{RCNSC11}.
Macroscopic Schr\"odinger Cat states are of particular interest 
in the context of the quantum-to-classical 
transition \cite{S07,Z03,C13}, tests of collapse models \cite{BG03,GPR90,R11,BLSSU14}
and the search for gravitational decoherence \cite{KHKRSJA12,B12,AL13,BGL08,SQ15,GU14}.
A major difficulty is the strong suppression of coherences since 
macroscopic systems cannot be isolated completely from their environment and
the emission and absorption of blackbody radiation washes out the interference pattern \cite{KHKRSJA12}.

The preparation of a spatial (macroscopic) superposition requires that the 
wavefunction of the composite object is split into two (or more) components.
This can be done, for example, with an apparatus which acts as a partially silvered mirror,
separating the incoming wave-packet into a reflected and a transmitted part.
In the following we will show that long-lived resonances can occur
due to the presence internal degrees of freedom of the composite object.
Furthermore, the wavepacket components can decohere partially when 
internal degrees of freedom are excited during the scattering process.
These effects are also present when the system is  
perfectly isolated from the environment.

During a scattering process of a composite object, it is likely that 
not all individual constituents interact directly with the mirror.
For example, Rutherford-scattering affects only the protons of 
$\alpha$-particles directly whereas the neutrons do not feel the Coulomb potential
of the heavy nuclei.
Since the protons and neutrons are bound to each other by nuclear forces,
the neutrons follow the trajectory of the protons. 
The situation that only a part of the composite object interacts 
with a scattering center is rather generic, the atom-light 
interaction being another example:
Only the dipole moment of the electron is affected by 
the electromagnetic field whereas the nucleus is not sensitive
to the wavelength of the photons.
%
%
%

A simple model Hamiltonian, which resembles the interaction of an single (center-of-mass)
degree of freedom with a mirror, is
\begin{align}
\hat{H}=\frac{\hat{p}^2}{2}+\hat{V}(\hat{x})\,,
\end{align}
where we assume that the potential has the form
$\hat{V}(\hat{x})=V_\mathrm{m}\delta(\hat{x})$.
An incoming plane wave with momentum $k$ will be partially transmitted and reflected, i.e.
\begin{align}
\psi_k(x)=\begin{cases}
     e^{ikx}+r e^{-ikx} & \text{for } x<0\\
     t e^{ikx} & \text{for } x>0\,,
     \end{cases}
\end{align}
where the amplitudes for transmission and 
reflection are determined by
\begin{align}
t=\frac{k}{k+iV_\mathrm{m}}\quad \text{and}\quad r=-\frac{iV_\mathrm{m}}{k+iV_\mathrm{m}}\,.
\end{align}

Choosing $k=V_\mathrm{m}$ we have
$|t|^2=|r|^2=1/2$ which mimics a half-silvered mirror.
The system exhibits a resonance for $ik_\mathrm{res}=V_\mathrm{m}>0$ and 
a bound state for $V_\mathrm{m}<0$.
Since we have $\mathrm{Re}(k_\mathrm{res})=0$, 
the dynamics of a wavepacket which is centered around 
a finite momentum value will not be affected by the resonance.
%
%
In the following we will see how the situation changes when the dynamics of internal 
degrees of freedom is taken into account.

\section{The model.} 

Consider two particles with unit mass and coordinates $x_1$ and $x_2$.
Both particles are tied to each other by a binding potential $\hat{V}_\mathrm{b}$
which depends only on the difference 
of the particle positions, $x_\mathrm{rel}=x_1-x_2$.
In general, we allow both particles to interact with a mirror
and introduce the scattering potentials $V_\mathrm{m}^{1/2}\delta(\hat{x}_{1/2})$.
The Hamiltonian which describes the scattering of this bound system 
is given by
\begin{align}\label{hamilt}
\hat{H}=\frac{\hat{p}_1^2}{2} +\frac{\hat{p}_2^2}{2}+V^1_\mathrm{m} \delta(\hat{x}_1)+V^2_\mathrm{m} \delta(\hat{x}_2)+
\hat{V}_\mathrm{b}(\hat{x}_1-\hat{x}_2)\,,
\end{align}

\begin{widetext}

\begin{figure}[t]
 \includegraphics[width=12cm]{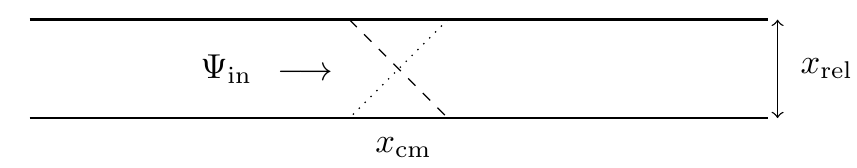}
 \caption{When the particles are bound to each other by a 
 hard wall potential, the system corresponds to a two-dimensional 
 waveguide. The scattering potential $V_\mathrm{m}^1\delta(\hat{x}_1)$
 is nonzero along the dotted line whereas $V_\mathrm{m}^2\delta(\hat{x}_2)$
 is nonzero along the dashed line.}\label{waveguide}
 \end{figure}
 
\end{widetext}
where $\hat{p}_{1,2}$ are the momentum operators of the individual particles.
%

It is useful to formulate the problem in terms of the center-of-mass coordinate 
$\hat{x}_\mathrm{cm}=\hat{x}_1+\hat{x}_2$ and the relative coordinate 
$\hat{x}_\mathrm{rel}=\hat{x}_1-\hat{x}_2$.
An arbitrary wave-packet is of the form 
\begin{align}\label{ansatz}
\Psi=\sum_n f_n(x_\mathrm{cm},t)\phi_n(x_\mathrm{rel})\,,
\end{align}
where the  $f_n$ are time-dependent functions 
which only depend center-of-mass coordinate and the $\phi_n$ fulfill eigenvalue equation
\begin{align}\label{eig}
-\frac{d^2 \phi_n}{d x_\mathrm{rel}^2}+V_\mathrm{b}(x_\mathrm{rel})\phi_n=
\epsilon_n \phi_n \,.
\end{align}
The choice of the binding potential determines how far the particles 
can separate from each other.
When we choose a hard wall potential, the system 
is related to a two-dimensional waveguide, see Fig.~\ref{waveguide}.

Inserting the ansatz (\ref{ansatz}) into the time-dependent Schr\"odinger equation, multiplying with $\phi_m(x_\mathrm{rel})$
and integrating over $x_\mathrm{rel}$ leads to a set of coupled equations,
\begin{align}\label{schroedinger}
i\partial_t f_n=-\frac{d^2 f_n}{d x_\mathrm{cm}^2}+\epsilon_n f_n+ \sum_m V_{nm}(x_\mathrm{cm})f_m\,. 
\end{align}
$V_{nm} $ is the effective potential which couples the functions $f_n$.
It can be expressed in terms of the  eigenfunctions $\phi_n$ of the binding potential, 
\begin{align}
V_{nm}(x_\mathrm{cm})=&2V^1_\mathrm{m}\phi_n(-x_\mathrm{cm})\phi_m(-x_\mathrm{cm})\nonumber\\
&+2V^2_\mathrm{m}\phi_n(x_\mathrm{cm})\phi_m(x_\mathrm{cm})\,.
 \end{align}
We assume that the wave-packet is initially of the form
\begin{align}\label{initial}
\Psi_\mathrm{in}=\left(\frac{2}{\pi\sigma^2}\right)^{1/4}\exp\left(i P x_\mathrm{cm}-
\frac{(x_{\mathrm{cm},0}-x_\mathrm{cm})^2}{\sigma^2}\right)\phi_0(x_\mathrm{rel})\,,
\end{align}
where the initial momentum of the center-of-mass coordinate is denoted with $P$ and the spread of the 
wave packet with $\sigma$.
The wave-packet is initially centered to the 
left of the mirror, $x_{\mathrm{cm},0}\rightarrow-\infty$, and follows 
a free evolution before it hits the mirror.
The ``internal'' degree of freedom is assumed to be in the ground state $\phi_0(x_\mathrm{rel})$.

\section{Harmonic coupling.}

A simple binding potential is the harmonic coupling 
$\hat{V}_\mathrm{b}=\Omega^2\hat{x}_\mathrm{rel}^2$ with stiffness $\Omega$. 
From the eigenvalue equation (\ref{eig}) we find that the 
energy levels of the internal degree of freedom are $\epsilon_n=2\Omega(n+1/2)$.
The effective coupling potential $V_{nm}$ is determined by the 
eigenfunctions of the harmonic oscillator, 
$\phi_n=((\Omega/\pi)^{1/4}/\sqrt{2^nn!})H_n(\sqrt{\Omega}x_\mathrm{rel})
\exp(-\Omega x_\mathrm{rel}^2/2 ) $. 

\subsection{Asymmetric scattering, $V^1_\mathrm{m}=0,V^2_\mathrm{m}>0$}

First, we discuss the situation 
when only one particle interacts directly mirror whereas particle 2 is only affected indirectly by the mirror via the 
 binding potential.
This corresponds to the choice
$V_\mathrm{m}^1=0$ and $V_\mathrm{m}^2>0$ in our model.
In oder to mimic a half-silvered mirror, we choose the parameter $V_\mathrm{m}^1$ and the initial momentum $P$
such that half of the 
wave-packet is transmitted and half of it is reflected,
In Fig.~\ref{dynamics} we show the reduced density matrix
$\rho_\mathrm{cm}~=~\int_{-\infty}^\infty dx_\mathrm{rel}\Psi^*(x_\mathrm{cm},x_\mathrm{rel})\Psi(x'_\mathrm{cm},x_\mathrm{rel})$
and the square of the modulus of the wavefunction at different 
times.
The wavepacket is localized before the scattering process, see Fig.~\ref{dynamics} (a).
Then, according to the sketch in Fig.~\ref{waveguide}, the scattering occurs at the 
line $x_2=0$ which can be seen in Fig.~\ref{dynamics} (b).
After the scattering process, the reflected and transmitted parts 
of the wavefunction separate from the mirror, see Fig.~\ref{dynamics} (c).

The scattering process is not 
adiabatic and higher modes $\phi_n$ ($n>0$)
become populated.
The initial product state (\ref{initial}) evolves to 
a general superposition (\ref{ansatz}).
In Fig.~\ref{modesHARM} (a) and Fig.~\ref{modesHARM} (b) we depict the time-evolution 
of 
$
p_{n,\mathrm{L}}=\int_{-\infty}^0 |f_n|^2 dx_\mathrm{cm}$
and
$p_{n,\mathrm{R}}=\int^{\infty}_0 |f_n|^2 dx_\mathrm{cm}$
which are the probabilities to find the internal degree of 
freedom in the state $\phi_n$ if the state 
is measured to the left or to the right from the mirror.
Since half of the wavepacket is transmitted and half of 
it is reflected, we have $\sum_n p_{n,\mathrm{L}}=\sum_n p_{n,\mathrm{R}}=0.5$ after the scattering.

The composite system is in a pure state
which allows us to characterize the excitation of the 
wave-packet by the entanglement-entropy, 
$S=-\mathrm{tr}(\hat{\rho}_\mathrm{cm}\ln\hat{\rho}_\mathrm{cm})$.
The entropy changes only during the scattering process
since the free time-evolution does not excite the internal 
degree of freedom, see Fig.~\ref{energyentropyPRD} (a),

\begin{widetext}
 
\begin{figure}[h]
\begin{center}
\includegraphics[width=18cm]{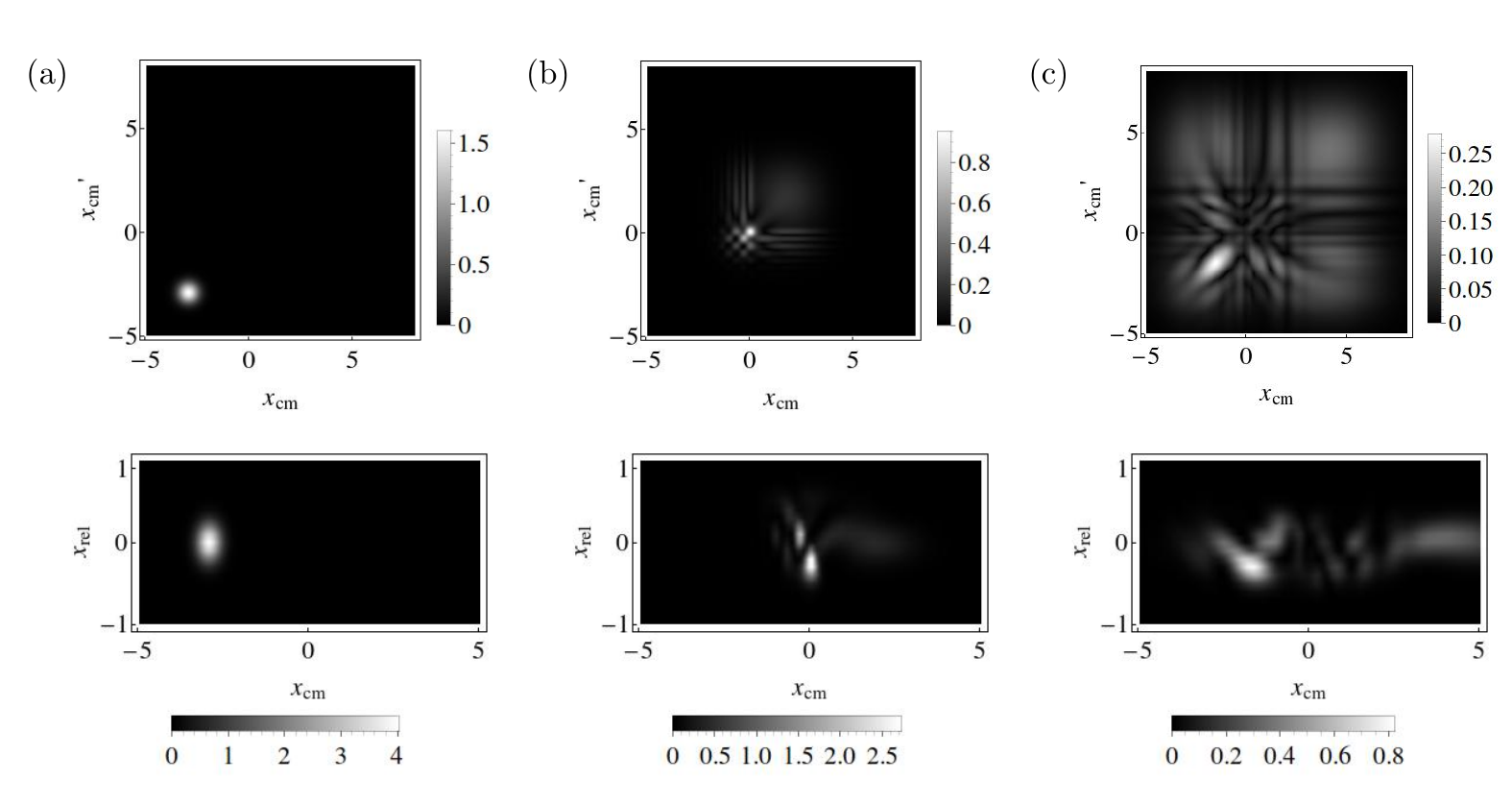}
\caption{Dynamics of the scattering process. \textit{Top:} Absolute value of the reduced density matrix $\rho_\mathrm{cm}$.
\textit{Bottom:} Absolute value of the wavefunction. (a)  Initial configuration 
of the wavepacket. 
(b) The scattering process occurs at the line $x_2=0$.
(c) The reflected part and the transmitted part of the wavefunction 
are separated from the mirror again.}\label{dynamics}
 \end{center}
 \end{figure}
 \vspace{-1cm}
\begin{figure}[h]
 \includegraphics[width=18cm]{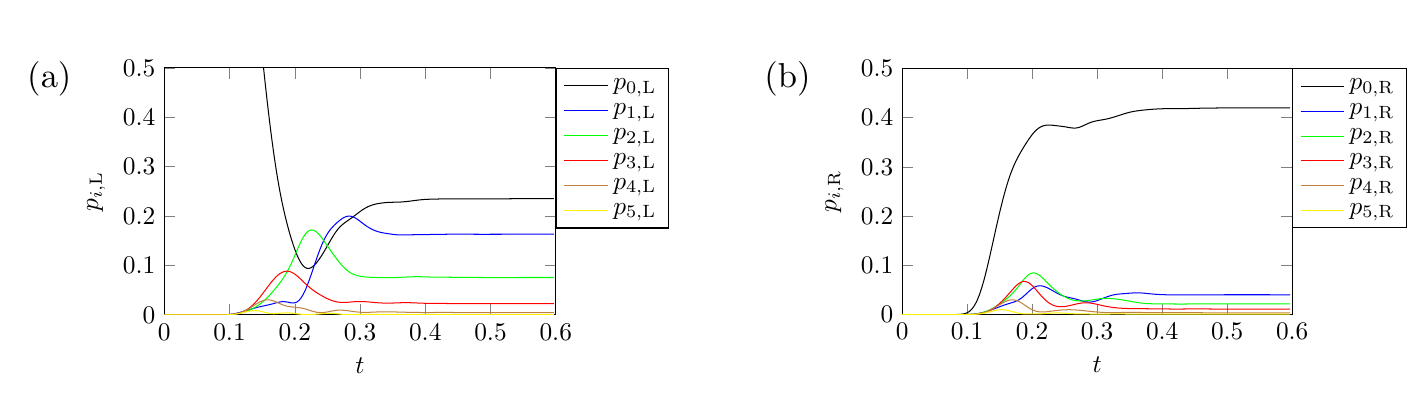}
\caption{
As explained in the text, the internal degree of freedom is 
more excited in the reflected part (a) than in the transmitted part (b).
In order to achieve that after the scattering process half of the 
wavepacket is transmitted and half of it is reflected, we choose
$P=10$, $V^\mathrm{2}_\mathrm{m}=11$, $\sigma=1/2$ and $\Omega=10$.}\label{modesHARM}
\end{figure}

 \end{widetext}

Furthermore, the energy of the center-of-mass coordinate and the energy of 
the internal degree of freedom is different for the reflected 
and transmitted part of the wave-packet, see Fig.~\ref{energyentropyPRD} (b).
The internal degree is more excited when the bound system 
is reflected: One particle is reflected by the mirror, the second 
one passed through and is pulled back through the mirror
due to the binding potential $V_\mathrm{b}$.
This leads to an increased excitation compared to the transmission where both
particles go through the mirror without the internal 
dynamics being involved.
This explains why the $p_{i>0,\mathrm{L}}$ (Fig.~\ref{modesHARM} (a)) are 
larger than the $p_{i>0,\mathrm{R}}$
(Fig.~\ref{modesHARM} (b)) for $t\rightarrow\infty$.
The reflected and transmitted parts of the center of mass motion become
disentangled.


\begin{widetext}

 \begin{figure}[h]
\includegraphics[width=18cm]{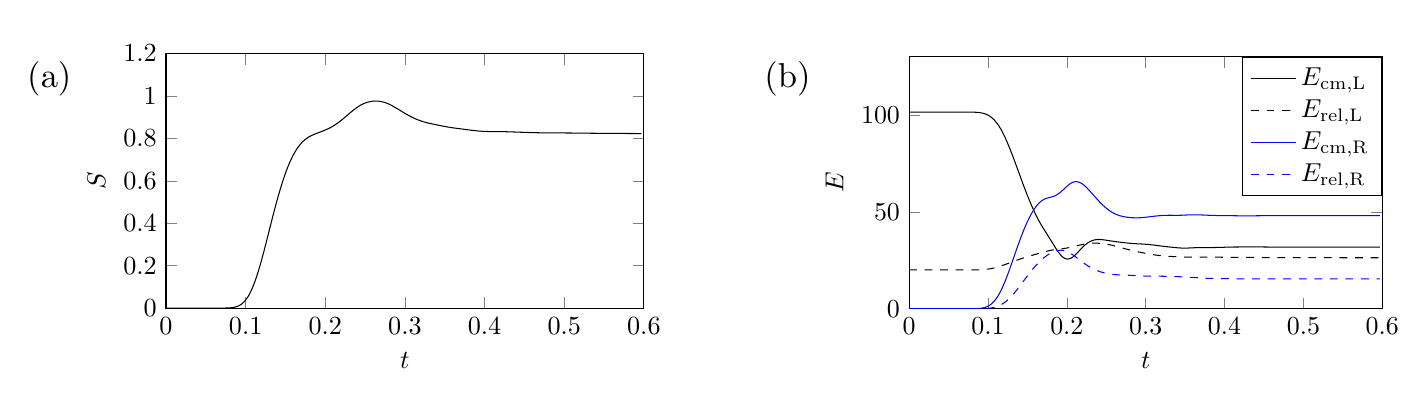}
\caption{(a) Time-evolution of the entanglement-entropy for asymmetric scattering. 
(b) Time-evolution of the center-of-mass energy 
$E_\mathrm{cm}=\langle \hat{p}_\mathrm{cm}^2\rangle$ and the internal energy
$E_\mathrm{rel}=\langle \hat{p}_\mathrm{rel}^2+\hat{V}_\mathrm{b}\rangle$
of the wave-packet to the left (L) and to the right (R) of the mirror.}\label{energyentropyPRD}
\end{figure}

\begin{figure}[h]
\begin{center}
\includegraphics[width=18cm]{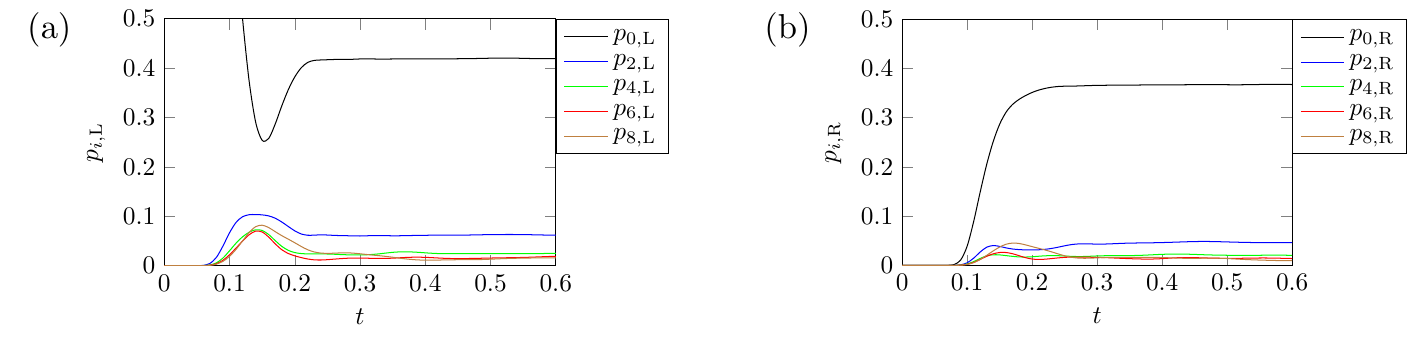}
\includegraphics[width=18cm]{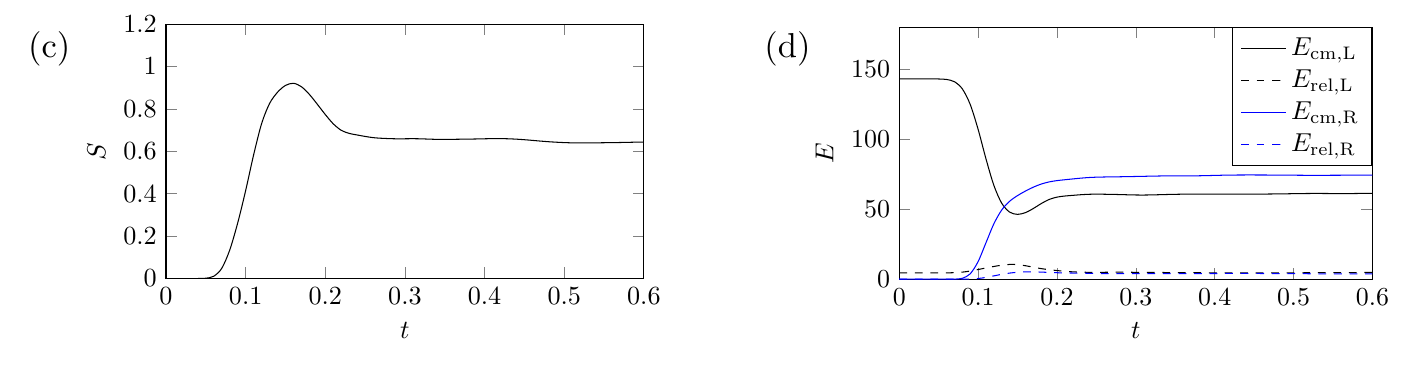}
\caption{Symmetric scattering of the two-particle bound state. 
(a) and (b) Time-evolution of the occupation probabilities for 
the modes of the internal degree of freedom to the left (L) and to the 
right (R) of the mirror.
Due to the symmetry of the scattering potential, the antisymmetric modes are not occupied.
In order to achieve that after the scattering process half of the 
wavepacket is transmitted and half of it is reflected, we choose
$P=12$, $V^\mathrm{1/2}_\mathrm{m}=15$, $\sigma=1/2$ and $\Omega=5$.
Time-evolution of the entanglement-entropy (c), the center-of-mass energy 
and internal energy (d).
}\label{modesdoublePRD}
\end{center}
 \end{figure} 
\vspace{5cm}
 \end{widetext}

\subsection{Symmetric scattering, $V^1_\mathrm{m}=V^2_\mathrm{m}>0$}

In the following we discuss the case 
when both particles interact directly with the mirror
with the same strength, i.e.~$V^1_\mathrm{m}=V^2_\mathrm{m}>0$.
Again, we choose the initial momentum and the value of $V^{1,2}_\mathrm{m}$
such that half of the wave-packet is transmitted and half of it 
is reflected.
The scattering potential does not mix the symmetric modes $\phi_n$, ($n=0,2,4,...$)
with the antisymmetric modes $\phi_n$, ($n=1,3,5,...$).
Therefore antisymmetric modes do not become excited when the 
system is initially in the ground state, see Fig.~\ref{modesdoublePRD}.
This in turn implies that the energy gap between ground 
state and first accessible excited state is $\epsilon_2-\epsilon_0=4\Omega$
rather than $\epsilon_1-\epsilon_0=2\Omega$.
As a consequence, the increase of entanglement-entropy 
is smaller for the symmetric scattering than it is 
for the asymmetric scattering, compare Fig.~\ref{energyentropyPRD} (a)
and Fig.~\ref{modesdoublePRD} (c).
Furthermore, the energy transfer from the center-of-mass motion 
to the relative motion is smaller, 
compare Fig.~\ref{energyentropyPRD}~(b)
and Fig.~\ref{modesdoublePRD}~(d).

\begin{figure}[t]
\begin{center}
\includegraphics[width=8.9cm]{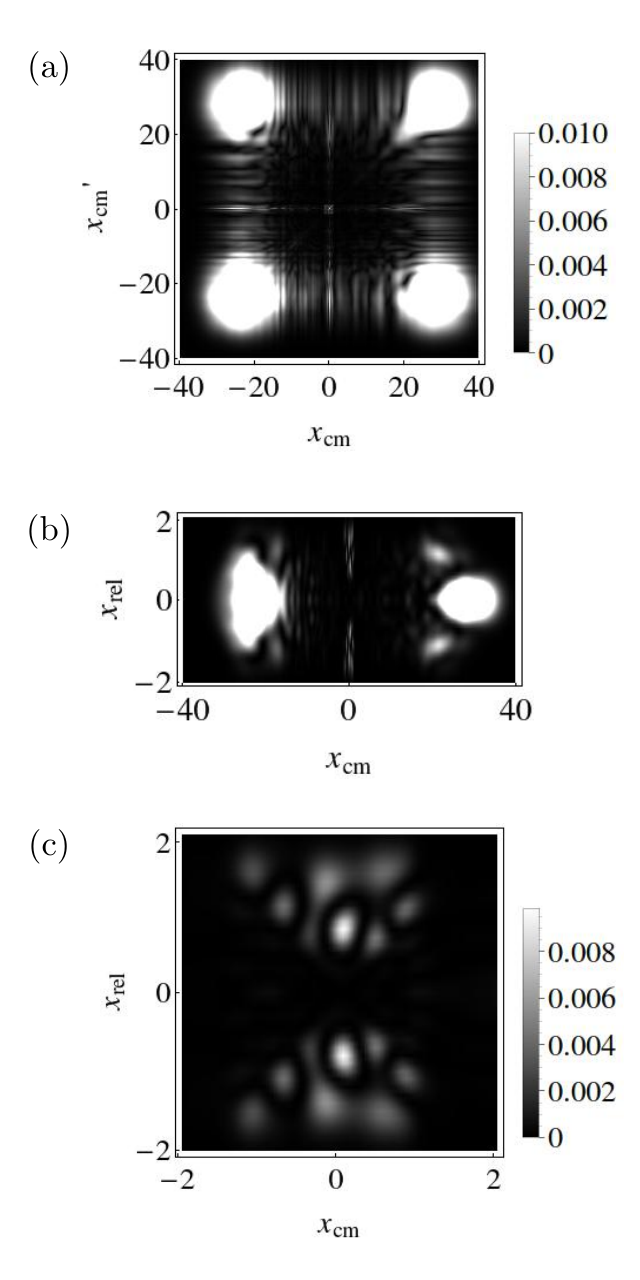}
\caption{Symmetric scattering of the bound state. (a) Reduced 
density matrix $\hat{\rho}_\mathrm{cm}$ after the scattering. A part of the 
wavefunction is trapped at the mirror. (b) Wavefunction after the scattering 
process. (c) The wavefunction is trapped in the triangles which are 
defined by the lines $x_1=0$ and $x_2=0$.
}\label{dynamicsdoublePRD}
\end{center}
 \end{figure}

An interesting feature of the symmetric scattering is that
it takes a rather long time before the occupation probabilities
become temporal constants.
This is due to the occurence of long-lived resonances.
Intuitively, this can be understood from the sketch
in Fig.~\ref{waveguide}:
When particle 1 is reflected at $x_1=0$, it pulls back particle 
2, then particle 2 is reflected at $x_2=0$ and particle 1 is pulled back.
This can happen several times until the wave-packet finally separates 
from the mirror.
As a result, a part of the wave-function is trapped in the ``triangles'' for a finite time.
Although most of the wave-function has separated from the 
mirror after the scattering, 
there is a non-vanishing probability to find the particles 
at the mirror.
The interference pattern of $\hat{\rho}_\mathrm{cm}$ after the scattering 
process is depicted in Fig.~\ref{dynamicsdoublePRD} (a) 
whereas the wave-function is shown in Fig.~\ref{dynamicsdoublePRD} (b).
In Fig.~\ref{dynamicsdoublePRD} (c) one can see how the 
wave-function is trapped in the triangles.

\section{Resonances.}

In general, resonances can be characterized as eigenvalues of the Hamiltonian which is 
subject to non-hermitian boundary conditions.
A simple example was given in the introduction: The scattering of a 
particle at a delta-function potential with positive prefactor $V_\mathrm{m}$ 
exhibits a resonance for the wavenumber $k_\mathrm{res}=-iV_\mathrm{m}$.
The corresponding wave-function is not in the Hermitian sector of the 
domain of the physical Hamiltonian and grows exponentially for large $|x|$, $\psi\sim e^{V_\mathrm{m}|x|}$.
In the following we use the complex scaling method to determine the position of the resonances \cite{M11,M98,M78,S78,RS78}:
The core of this method is a transformation 
of the Hamiltonian such that the resonant 
eigenfunctions become square integrable.
For the problem at hand, this can be achieved
by transforming the the center-of-mass coordinate according to 
$x_\mathrm{cm}\rightarrow e^{i\theta}x_\mathrm{cm}$.
When the angle $\theta$ is large enough, the 
eigenfunction of a particular resonance
decreases exponentially for $|x_\mathrm{cm}|\rightarrow\pm\infty$.
Thus, the complex energy $E_\mathrm{res}$
can be determined from the
eigenvalue problem for the transformed Hamiltonian.
The value of $E_\mathrm{res}$ is locally independent 
on $ \theta$ and appears as isolated 
point in the spectrum, see Appendix.
The life-time $\tau$ of a resonance is determined 
by the imaginary part of the resonant energy 
via $\tau=-1/(2\mathrm{Im}E_\mathrm{res})$.
Thus, the closer the resonant energy is to 
the real axis, the longer is the decay time.

For asymmetric scattering, we find that the resonances are far away from the real axis, 
see Fig.~\ref{resonancesPRD} (a).
However, when both particles interact with the 
mirror, i.e.~$V_\mathrm{m}^1=V_\mathrm{m}^2>0$, 
the resonances can be long-lived, see Fig.~\ref{resonancesPRD} (b).
The effect on the dynamics becomes apparent 
from the amplitude of the wavefunction 
at the mirror:
For asymmetric scattering, the amplitude at the mirror vanishes 
immediately (Fig.~\ref{dynamics}) whereas the decrease of the amplitude 
during symmetric scattering is much slower (Fig.~\ref{dynamicsdoublePRD}).
In Fig.~\ref{wavepacketsPRD} (a), we show
the amplitude for symmetric scattering at various times.
Even when the parameters of the model are choosen 
such that most of the bound system is reflected 
(corresponding to a totally reflecting mirror),
a part of the wave-function is trapped temporarily at the repulsive 
scattering potential, see
Fig.~\ref{wavepacketsPRD}~(b).

\newpage
\begin{widetext}

\begin{figure}[h]
\begin{center}
\includegraphics[width=18cm]{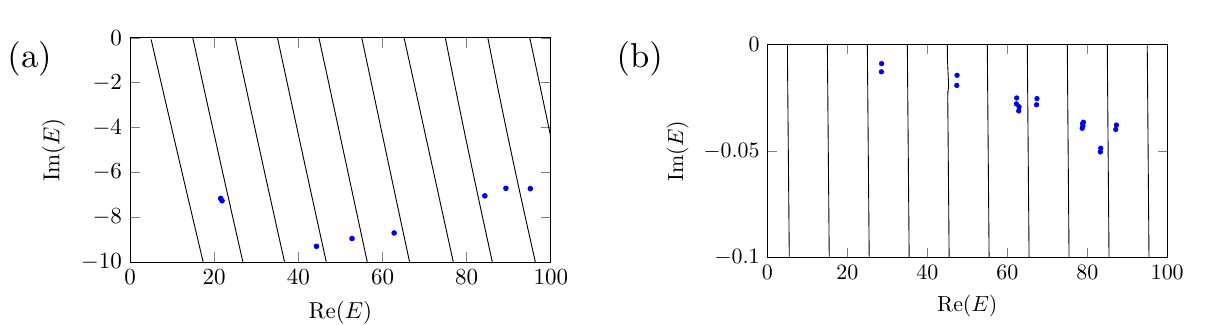}
\caption{Spectrum of the Hamiltonian after analytical 
continuation. The lines denote the continuous part of the spectrum 
and the blue dots are the resonances. (a) Asymmetric scattering: The parameters are 
$\Omega=~5$, $V_\mathrm{m}^1=0$, $V_\mathrm{m}^2=30$ and $\theta=0.35$. 
The resonances are short-lived when 
only particle interact directly with the mirror.
(b)~Symmetric scattering: The parameters are 
$\Omega=5$, $V_\mathrm{m}^1=V_\mathrm{m}^2=30$ and $\theta=0.1$. 
The resonances can be long-lived when 
both particles interact directly with the mirror.
Note that the inverse life-times $\mathrm{Im}(E)$ for asymmetric scattering (a) 
and for symmetric scattering (b) differ by a factor of 100 for the same choice 
of parameters.
The resonance which is closest to the real axis has an inverse lifetime of 
 $|\mathrm{Im}(E_\mathrm{res})|=0.0089$.}\label{resonancesPRD}
\end{center}
\end{figure}

\begin{figure}[h]
\begin{center}
\includegraphics[width=18cm]{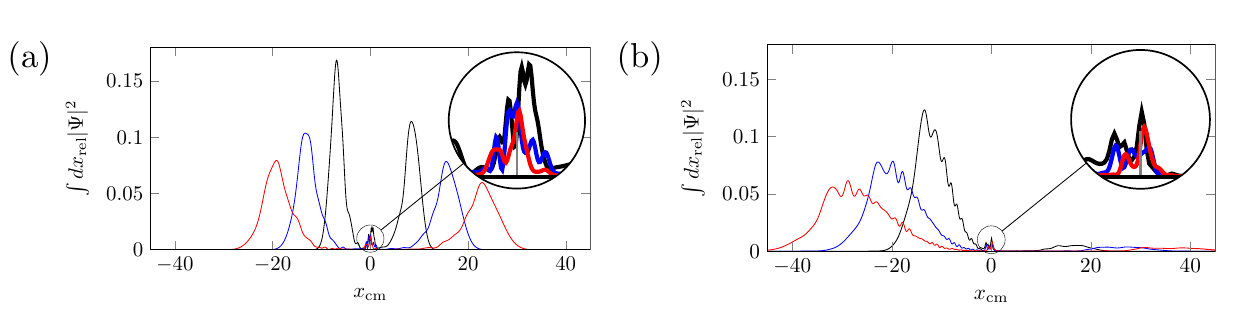}
\caption{(a) Wave-function for $t=0.5 $ (black), $t=0.8 $ (blue), and $t=1.1 $ (red). 
 The inset shows the part of the wave-packet
 which is trapped at the mirror. The parameters are the 
 same as in Fig.~\ref{modesdoublePRD}.
 (b) Wave-function for $t=1 $ (black), $t=1.5 $ (blue), and $t=2 $ (red). We selected
 the parameters $V_\mathrm{m}^{1,2}=20$, $\Omega=5$, $P=8$, and $\sigma=1/2$.}\label{wavepacketsPRD}
\end{center}
\end{figure}

\end{widetext}

\section{WKB analysis}

It is possible to analyze the Schr\"odinger equation 
(\ref{schroedinger}) within a WKB approximation
which will give a qualitative understanding for the 
occurence of the resonances.
For this, we turn to the time-independent version 
of (\ref{schroedinger}) using the ansatz $\mathbf{f}=\mathbf{f}_E e^{-iE t}$.
After reintroducing $\hbar$, the time-independent Schr\"odinger equation has the form
\begin{align}
E\,\mathbf{f}_E=-\hbar^2\frac{d \mathbf{f}_E}{dx^2_\mathrm{cm}}+(\hat\epsilon+\hat V)\mathbf{f}_E\,. 
\end{align}
We choose the ansatz
\begin{align}
\mathbf{f}_E=\hat M e^{i\frac{\hat S_0}{\hbar}}\hat A \mathbf{v}\,,
\end{align}
where $\mathbf{v}=(0,...0,1,0,...,0)$ is a unit vector 
and $\hat M$, $\hat A$ and $\hat S_0$ are matrices.
Up order unity we obtain a first-order differential 
equation for $\hat S_0$,
\begin{align}
\left(\frac{d\hat S_0}{dx_\mathrm{cm}}\right)^2+\hat M^{-1}(\hat \epsilon+\hat V)\hat M=E.
\end{align}
If we demand that the matrix $\hat S_0$ is diagonal, the matrix
$\hat M^{-1}(\hat \epsilon+\hat V)\hat M=\hat V_D$ has to be of diagonal form which 
in turn determines $\hat M$.
Up to order $\hbar$, we obtain as a condition for $\hat A$
\begin{align}\label{hbar}
0=&2 e^{-i\frac{\hat S_0}{\hbar}}\left(\frac{d\hat S_0}{dx_\mathrm{cm}}\right)^{-1}
\hat M^{-1}\frac{d\hat M}{dx_\mathrm{cm}}\left(\frac{d\hat S_0}{dx_\mathrm{cm}}\right)e^{i\frac{\hat S_0}{\hbar}}\nonumber\\
&+2\left(\frac{d\hat A}{dx_\mathrm{cm}}\right)\hat A^{-1}+\frac{d^2\hat S_0}{dx_\mathrm{cm}^2}\,.
\end{align}

\begin{widetext}

\begin{figure}[t]
\centering
\includegraphics[width=18cm]{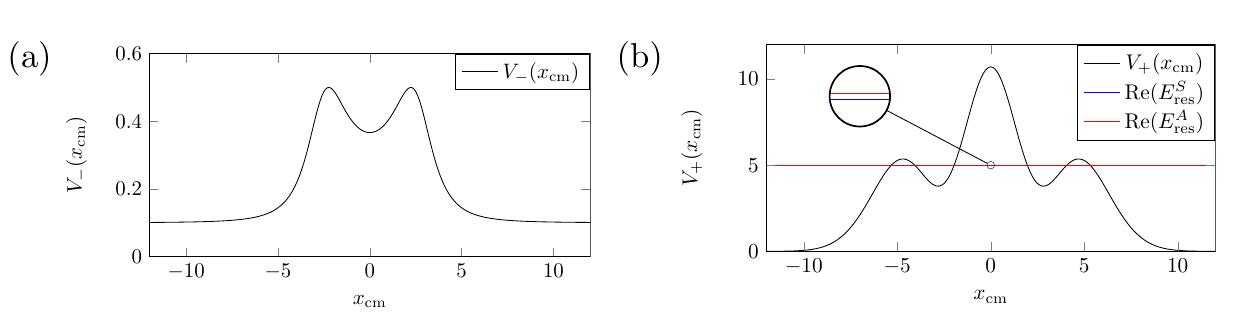}
\caption{Effective potentials for $\Omega=0.1$ and $V=10$.
(a) $V_-$ has no resonance since the local minimum is too shallow.\\
(b) $V_+$ has two resonances which are related to a symmetric 
and an antisymmetric state. The inset shows the energy 
splitting of the states.}\label{potentials}
\end{figure}

 \end{widetext}

If the potential matrix were spatially constant, the first term in
equation (\ref{hbar}) would vanish.
For simplicity we assume that the potential matrix $\hat V$ variies slowly 
such that we can neglect $d\hat M/dx_\mathrm{cm}$.
The WKB solution takes then the form
\begin{align}
\mathbf{f}^\mathrm{WKB}_E(x_\mathrm{cm})&=\hat M(x_\mathrm{cm}) \mathbf{g}^\mathrm{WKB}_E(x_\mathrm{cm})
\end{align}
with
\begin{align}
\mathbf{g}^\mathrm{WKB}_E(x_\mathrm{cm})=
\frac{\exp\left(\pm i\int_0^{x_\mathrm{cm}}dx_\mathrm{cm}'\sqrt{E-\hat V_D(x_\mathrm{cm}')}\right)}{\sqrt{E-\hat V_D(x_\mathrm{cm})}}
\mathbf{v}\,.
\end{align}
Here we assumed that the the energy is larger than the entries of the potential matrix.
Otherwise one has exponential increasing or decreasing WKB-solutions.

In the following, we limit our considerations 
to the modes $n=0$ and $n=2$ and assume a
symmetric scattering process with
$V_\mathrm{m}^1=V_\mathrm{m}^2=V>0$.
Therefore, we assume that the excitations related to 
$n=4,6,...$ can be disregarded.
Within the order of approximation, we have two modes
\begin{align}
\mathbf{g}^\mathrm{WKB}_{E,-}(x_\mathrm{cm})=g^\mathrm{WKB}_- (x_\mathrm{cm})
\begin{pmatrix}
0\\1 
\end{pmatrix}
\end{align}
and 
\begin{align}
\mathbf{g}^\mathrm{WKB}_{E,+}(x_\mathrm{cm})=g^\mathrm{WKB}_+ (x_\mathrm{cm})
\begin{pmatrix}
1\\0 
\end{pmatrix}
\end{align}
which are WKB-solutions of the eigenvalue equations
\begin{align}
-\frac{d^2 g_\pm}{dx_\mathrm{cm}^2}+V_\pm(x_\mathrm{cm}) g_\pm=Eg_\pm\,.
\end{align}
The potentials are given by
\begin{align}
V_\pm&= \frac{1}{2}\bigg[\epsilon_0+\epsilon_2+4 V(\phi_0^2+\phi_2^2)\nonumber\\
&\pm\sqrt{[\epsilon_0-\epsilon_2+4V(\phi_0^2-\phi_2^2)]^2+
64V^2\phi_0^2\phi_2^2}\bigg]\,.
\end{align}
For positive $V$, we have $V_\pm(x_\mathrm{cm})>V_\pm(x_\mathrm{cm}\rightarrow\pm\infty)$
such that bound states are not possible but resonances can appear.
Indeed, $V_-$ has one local minimum whereas $V_+$ has two local minima, 
see Fig.~\ref{potentials}.
If the scattering potential is much larger than the 
internal level spacing, $V\gg\Omega^{1/2}$, the potentials can be approximated 
by
\begin{align}
V_-(x_\mathrm{cm})=\frac{\Omega(11-4\Omega x_\mathrm{cm}^2+4\Omega^2 x_\mathrm{cm}^4)}
{3-4\Omega x_\mathrm{cm}^2+4\Omega^2 x_\mathrm{cm}^4}
\end{align}
and
\begin{align}
V_+(x_\mathrm{cm})=\frac{2\sqrt{\Omega}Ve^{-\Omega x_\mathrm{cm}^2}}{\sqrt{\pi}}
(3-4\Omega x_\mathrm{cm}^2+4\Omega^2 x_\mathrm{cm}^4)\,.
\end{align}
For the potential $V_-$ (see Fig.~\ref{potentials} (a)), we find that the local minimum 
is too shallow to support a resonance.
In contrary, the potential $V_+$ (see Fig.~\ref{potentials} (a)) 
supports for $V\gg\Omega^{1/2}$
two resonances which correspond to a symmetric and an 
anti-symmetric state.
The energies are given by
\begin{align}
E^{S/A}_\mathrm{res}=V_\mathrm{min}+\omega\pm \frac{\omega}{\pi}e^{-\alpha}-\frac{i \omega}{\pi}e^{-2\beta}\,,
\end{align}
where $V_\mathrm{min}\approx1.20\,\Omega^{1/2}V$ and $\omega\approx 2.14\, \Omega^{3/4}V^{1/2}$ 
are the value of the potential $V_+$ at the local minima 
and the stiffness, respectively.
Furthermore, the exponent $\alpha\approx-1.20+1.22\, \Omega^{-1/4}V^{1/2}$ determines the tunnel coupling 
between the minima and the exponent $\beta\approx -1.91+0.45\,\Omega^{-1/4}V^{1/2}$
determines the decay rate of the resonance.

\section{Conclusion.}

We have shown with a simple model that resonances 
can occur in the scattering process of a composite system.
In particular we studied the splitting of a wave-packet 
by a potential which mimics a partially silvered mirror.
Depending on the particular system-mirror we found long-living
and short-living resonances.
If both particles interact with the mirror, it is possible 
that the composite system is trapped at the mirror for a
finite time.
We want to emphasize that these resonances occur 
due to the interaction between a purely \textit{repulsive} potential
(mirror) and the internal attractive potential.
In contrast, resonances 
and even bound states would not be surprising
for an attractive mirror potential.

When internal degrees of freedom become excited during the 
scattering process, partial which-path-information can be obtained
since the modes of the reflected and the transmitted wave-packet are populated differently.
Our findings should be of importance in the context of 
state preparation of mesoscopic objects 
or double-slit experiments with composite systems.
In particular, the excitations of internal degrees of freedom 
have to be taken into account when the superposition
principle is tested on a macroscopic scale, e.g. with optical trapped microspheres \cite{KHKRSJA12,AKM14}.

\section*{Acknowledgements.}

One of the authors (F. Q.) would like to thank R.~Froese,
M.~Choptuik for helpful comments.
This work is supported by the Templeton foundation
(grant number JTF 36838) and NSERC.

\newpage
\section*{Appendix}

In the following we give a brief introduction 
into the method of complex scaling which allows to 
determine the position of resonances.

Resonances result from imposing outgoing boundary conditions 
on the eigenfunctions of a time-independent Hamiltonian.
For our discussion, we consider the Hamiltonian of a particle 
which moves in a potential $V(\hat{x})$.
The Hamiltonian is given by
\begin{align}
\hat{H}=\hat{p}^2+\hat{V}(\hat{x}) 
\end{align}
and the asymptotic form of a general solution to the time-independent 
Schr\"odinger equation takes the form
\begin{align}
\psi_k(x\rightarrow+\infty)&=A_+(k)e^{-ik x}+B_+(k)e^{+ik x}\\
\psi_k(x\rightarrow-\infty)&=A_-(k)e^{+ik x}+B_-(k)e^{-ik x}\,.
\end{align}
Here we assumed that the potential vanishes for $x\rightarrow \pm \infty$ and $k$ is the asymptotic momentum of the particle.
When the amplitude of the incoming wave, $A_\pm$, vanishes at $k=k_\mathrm{res}$, the 
system exhibits a bound state (for $\mathrm{Re}(\mathrm{res})>0$ and $\mathrm{Im}(\mathrm{res})=0$)
or a resonance (for $\mathrm{Re}(\mathrm{res})>0$ and $\mathrm{Im}(\mathrm{res})<0$).
The imaginary part of the corresponding energy, $E_\mathrm{res}=k_\mathrm{res}^2$ is the 
lifetime of the resonance.
The complex wave vector is given by $k_\mathrm{res}=|k_\mathrm{res}|e^{-i\phi}$, where
\begin{align}
\phi=-\frac{1}{2}\mathrm{arctan}\left(\frac{\mathrm{Im}(E_\mathrm{res})}{\mathrm{Re}(E_\mathrm{res})}\right)
\end{align}
lies in the range $0<\phi<\pi/2$ for a resonant eigenfunction.
Asymptotically, the eigenfunctions adopt the form
\begin{align}
\psi_{k_\mathrm{res}}(x\rightarrow+\infty)&=B_+(k_\mathrm{res})e^{i|k_\mathrm{res}|\cos(\phi)x}e^{|k_\mathrm{res}|\sin(\phi)x}\label{asym1}\\ 
\psi_{k_\mathrm{res}}(x\rightarrow-\infty)&=B_-(k_\mathrm{res})e^{-i|k_\mathrm{res}|\cos(\phi)x}e^{-|k_\mathrm{res}|\sin(\phi)x}\label{asym2}\,.
\end{align}
Since the functions diverge for $x\rightarrow\pm\infty$ they are not eigenfunctions of an hermitian operator.
Complex scaling is based on a scaling transformation
of the time-independent Schr\"odinger equation such that the (transformed) 
resonant eigenfunction becomes square-integrable.
Then the resonant eigenenergies can be determined from 
an eigenvalue equation of the transformed (non-hermitian) Hamiltonian.
A general scaling operator is of the form
\begin{align}
\hat{S}=\sqrt{\eta} e^{\ln\eta x\frac{\partial}{\partial x}}
\end{align}
and transforms a wave-function according to
\begin{align}
\hat{S}\psi(x)=\sqrt{\eta}\psi(\eta x)\,.
\end{align}
Choosing $\eta=e^{i\theta}$, the coordinate is rotated into the complex plane.
The Hamiltonian is transformed according to
\begin{align}
\hat{H}(\theta)=\hat{S}\hat{H}\hat{S}^{-1}=\hat{p}^2 e^{-2i\theta}+\hat{V}(\hat{x}e^{i\theta})\,,
\end{align}
where we assumed that the potential can be analytically continued into the complex plane.
When the system has no energy threshold, the energies of the continuum states are 
$E_\mathrm{cont}=k^2$.
The states themselves are combinations of ingoing
and outgoing waves,
\begin{align}
\psi^\mathrm{cont}_k(x)=A(k)^{-ikx}+B(k)^{ikx}
\end{align}
and will be transformed under complex scaling to
\begin{align}
\psi^\mathrm{cont}_k(xe^{i\theta})=A(k)^{-ikxe^{i\theta}}+B(k)^{ikxe^{i\theta}}\,.
\end{align}
The only solutions which do not diverge have the wave vectors $k=|k|e^{-i\theta}$.
Thus, the continuum energies are rotated into the complex plane 
according to $E_\mathrm{cont}(\theta)=E_\mathrm{cont}e^{-2i\theta}$.
Since we assumed that the system has no energy threshold, the 
branch point of the rotated continuum is the origin of the 
complex energy plane.

A resonant wavefunction with the asymptotics (\ref{asym1}) and (\ref{asym2})
becomes square-integrable when 
\begin{align}\label{crit}
\theta\geq \phi\,.
\end{align}
The corresponding eigenvalue $E_\mathrm{res}$ of the analytical 
continued (non-hermitian) Hamiltonian $H(\theta)$ appears as \textit{isolated} point in the spectrum
when the condition (\ref{crit}) is satisfied.
It can be shown that $E_\mathrm{res}$ is locally independent of $\theta$
although the corresponding transformed eigenfunction $\psi_{k_\mathrm{res}}(xe^{i\theta}) $ 
varies with $\theta$.

For the analysis of the two-particle bound system, it is necessary to 
generalize the discussion.
Due to the internal structure of the bound system, there are infinitely many coupled mode functions $f_n$.
Each function $f_n$ corresponds to an energy threshold $\epsilon_n$
and has the asymptotic form
\begin{align}
f_ {n,k}(x\rightarrow+\infty)=&A_{n,+}(k)e^{-i\sqrt{k^2-\epsilon_n} x}\nonumber\\
&+B_{n,+}(k)e^{+i\sqrt{k^2-\epsilon_n} x}\\
f_ {n,k}(x\rightarrow-\infty)=&A_{n,-}(k)e^{+i\sqrt{k^2-\epsilon_n} x}\nonumber\\
&+B_{n,-}(k)e^{-i\sqrt{k^2-\epsilon_n} x}\,.
\end{align}
At a resonance, at least one of the mode functions satisfies 
the outgoing boundary condition
\begin{align}
f_{n,k_\mathrm{res}}(x_\mathrm{cm})\sim
\begin{cases}
e^{i \sqrt{k_\mathrm{res}^2-\epsilon_n} x_\mathrm{cm}}  &\text{ for } x\rightarrow\infty\\
e^{-i \sqrt{k_\mathrm{res}^2-\epsilon_n} x_\mathrm{cm}} &\text{ for }  x\rightarrow-\infty\,.
\end{cases} 
\end{align}
The asymptotic form of this component of the wavefunction reads
\begin{align}
f_{n,k_\mathrm{res}}(x_\mathrm{cm}\rightarrow\infty)&\sim e^{i a_n x_\mathrm{cm}}e^{b_n x_\mathrm{cm}}\rightarrow\infty
\end{align}
which diverges for large $x_\mathrm{cm}$ since
\begin{align}
a_n&=|E_\mathrm{res}-\epsilon_n|^{1/2}\cos(\phi_n)\\
b_n&=|E_\mathrm{res}-\epsilon_n|^{1/2}\sin(\phi_n)>0\\
\phi_n&=\frac{1}{2}\mathrm{arctan}\left(\frac{\mathrm{Im}(E_\mathrm{res})}{\mathrm{Re}(E_\mathrm{res})-\epsilon_n}\right)\,.
\end{align}
Acting with a complex-scaling operator on the wavefunction which transforms 
$x_\mathrm{cm}\rightarrow e^{i\theta}x_\mathrm{cm}$, we find that the
wavefunction becomes square integrable for angles $\theta\geq\phi_n$.
The continuum energies are rotated into the complex 
and start at the branch points $\epsilon_n$,
\begin{align}
E_\mathrm{cont}(\theta)=\epsilon_n+(E_\mathrm{cont}-\epsilon_n)e^{-2i\theta}\,. 
\end{align}

\end{document}